\documentclass[
	aps,
	prb,
	superscriptaddress, 
	preprintnumbers,
	amsmath,
	amssymb,
	longbibliography,
	twocolumn
	]{revtex4-2}
 \newcommand{\figurewidth}{.9\columnwidth}
\usepackage[utf8]{inputenc}
\usepackage[T1]{fontenc}

\usepackage{graphicx}
\usepackage{siunitx}	
\usepackage{amssymb}
\usepackage{datetime}
\usepackage[super]{nth}
\usepackage{pdfpages}
\usepackage{hyperref}
\usepackage{xcolor}
\usepackage{pgffor}
\hypersetup{
    colorlinks,
    linkcolor={red!50!black},
    citecolor={blue!50!black},
    urlcolor={blue!80!black}
}

\usepackage{todonotes}

\makeatletter
\AtBeginDocument{\let\LS@rot\@undefined}
\makeatother

\newcommand{\TiSe}{1T-TiSe$_2$}
\newcommand{\NbSe}{\texorpdfstring{2H-NbSe$_2$}{NbSe2}}

\newcommand{\EF}{\ensuremath{{E}_{\text F}}}

\newcommand{\TCDW}{\ensuremath{T_{\text{CDW}}}}
\newcommand{\DeltaCDW}{\ensuremath{\Delta_{\text{CDW}}}}
\newcommand{\PCDW}{\ensuremath{P_{\text{CDW}}}}

\newcommand{\Hcii}{\ensuremath{H_{\text{c2}}}}
\newcommand{\Tc}{\ensuremath{T_{\text c}}}

\newcommand{\Pc}{\ensuremath{P_{\text c}}}
\newcommand{\vF}{\ensuremath{v_{\text F}}}
\newcommand{\GPa}{\giga\pascal}

\newcommand{\RH}{\ensuremath{R_{\text H}}}
\newcommand{\Vl}{\ensuremath{V_{\text l}}}
\newcommand{\Vt}{\ensuremath{V_{\text t}}}

\newcommand{\dd}{\text{d}}										

\renewcommand{\vec}[1]{\boldsymbol{#1}}

\begin{document}

%
%


\title{
Absence of superconducting dome at the charge-density-wave quantum phase transition in \NbSe}

\author{Owen Moulding}
\affiliation{HH Wills Laboratory, University of Bristol, Bristol, BS8 1TL, UK}

\author{Israel Osmond}
\affiliation{HH Wills Laboratory, University of Bristol, Bristol, BS8 1TL, UK}

\author{Felix Flicker}
\affiliation{Rudolf Peierls Centre for Theoretical Physics, University of Oxford, Department of Physics,
Clarendon Laboratory, Parks Road, Oxford, OX1 3PU, United Kingdom}

\author{Takaki Muramatsu}
\affiliation{HH Wills Laboratory, University of Bristol, Bristol, BS8 1TL, UK}

\author{Sven Friedemann}
\email{Sven.Friedemann@bristol.ac.uk}
\affiliation{HH Wills Laboratory, University of Bristol, Bristol, BS8 1TL, UK}

\date{\today}


\begin{abstract}
	Superconductivity is often found in a dome around quantum critical points, i.e. \nth{2}-order quantum phase transitions. Here, we show that an enhancement of superconductivity is avoided at the critical pressure of the charge-density-wave (CDW) state in \NbSe. 
We present comprehensive high-pressure Hall effect and magnetic susceptibility measurements of the CDW and superconducting state in \NbSe. 
Initially, the \nth{2}-order CDW transition is suppressed smoothly but it drops to zero abruptly at $\PCDW = \SI{4.4}{\GPa}$ thus indicating a change to \nth{1} order whilst
the superconducting transition temperature \Tc\ rises continuously up to \PCDW\ but is constant above. 
The putative \nth{1}-order nature of the CDW transition is suggested as the cause for the absence of a superconducting dome at \PCDW.
Indeed, we show that the suppression of the superconducting state at low pressures is due to the loss of density of states inside the CDW phase whilst the initial suppression of the CDW state is accounted for by the stiffening of the underlying bare phonon mode. 

\end{abstract}

\maketitle

%
%
\section{Introduction}

The interplay of competing orders is of fundamental and practical interest \cite{Nakano2012,Baek2014,Friedemann2018,Chang2012}. 
Controlled switching between phases promises new applications in data storage and sensing \cite{Nakano2012}.
 On a fundamental level, understanding the interplay between ground states  provides important insight into the mechanism underlying each ground state and can reveal new phenomena at the border of ordered phases \cite{Mathur1998,Brando2016,Loehneysen2007}. For instance, a large body of work focuses on the interplay of superconductivity and charge order in cuprate high-temperature superconductors \cite{Frano2020}. 

With both superconductivity and charge-density-wave (CDW) order stabilised by the opening of a gap on (parts of) the Fermi surface, a mutual competition between the two states has been anticipated since early studies \cite{Bilbro1976}. As an alternative, superconductivity in a dome around quantum critical points was suggested to be promoted  by quantum fluctuations of the ordered state with prominent examples in heavy-fermion antiferromagnets \cite{Mathur1998}, CDW systems \cite{Morosan2006a}, and the CDW and pseudogap order in cuprate superconductors \cite{Ramshaw2015}. In addition to  competition and  promotion, superconductivity and charge order can coexist for instance by opening a gap on different parts of the Fermi surface; \NbSe\ is a prototypical material hosting both CDW order and superconductivity. However, fundamental questions about the interplay of the CDW and superconductivity remain open.

The interplay of CDW order and superconductivity in \NbSe\ remains disputed \cite{Kiss2007,Borisenko2009}.
CDW order sets in at $\TCDW\sim\SI{33}{\kelvin}$ while superconductivity is present below $\Tc=\SI{7.1}{\kelvin}$ at ambient pressure \cite{Moncton1975,Yokoya2001,Wilson1969,Wieteska2019}. Superconductivity opens  gaps of different sizes on most of the Fermi surface, while the CDW opens a gap on small parts of the zone-corner niobium-derived Fermi surface sheets only \cite{Rahn2012, Borisenko2009, Yokoya2001,Valla2004,Kiss2007,Fletcher2007,Galvis2018,Rahn2012}.
The separation of the CDW and superconducting gaps in $k$-space was interpreted as a hallmark for coexistence of the two ordered states. In addition, some studies suggested that superconductivity is boosted by the static CDW order \cite{Kiss2007} while further studies suggested a promotion of superconductivity by the soft modes present at the quantum critical point of the CDW order \cite{Feng2012,Suderow2005}. Finally, some studies suggested a bidirectional competition for density of states between the CDW and superconductivity \cite{Suderow2005,Dalrymple1984,Cho2018,Borisenko2009}.  Here, we use comprehensive high-pressure tuning of the CDW and superconducting states 
to reveal the absence of a superconducting dome ruling out a promotion of superconductivity by soft modes of the CDW. Rather, we show very clearly that superconductivity is reduced inside the CDW phase because of the loss of density of states. At the same time, we find indications of a \nth{1}-order CDW transition which is likely to be the reason for the absence of a dome-shaped superconducting phase at the critical pressure of the CDW.

%
%

\section*{Experimental Methods}
\subsection{Samples}
\label{subsec:Methods:Samples}
\NbSe\ samples were grown by J. A. Wilson \cite{Wilson1969} using the vapour transport method and have a high residual resistivity ratio, $\rho(T=\SI{300}{\kelvin})/\rho(T=\SI{9}{\kelvin})> 60$, confirming the good crystal quality. Samples were cut with a scalpel. 
Lateral sample dimensions have been obtained with an optical microscope at ambient pressure (see inset in \autoref{fig:SC}(a) of the main text), the sample thickness, $t$, was estimated from the sample mass and the lateral dimensions using the known density of \NbSe. The associated uncertainty of \SI{10}{\percent} results in a systematic relative uncertainty of the Hall coefficient of the same amount. 
The magnetic field was applied along the crystallographic $c$ direction. 

\subsection{High-Pressure Measurements}
\label{subsec:Methods:Pressure}
High-pressure measurements used moissanite anvils cells with a culet size of \SI{800}{\micro\meter} for both the electrical and magnetic measurements. 
Both types of measurement used  metallic gaskets which were prepared by indenting 450\si{\mu m} thick BeCu to approximately 60\si{\mu m} followed by drilling a 450\si{\mu m} hole.

Pressure was determined at room temperature by ruby fluorescence, with multiple ruby flakes placed within the sample chamber as a manometer. The uncertainty of the pressure is taken as the standard deviation between pressure estimates from rubies across the sample chamber, both before and after a measurement.
A comparison with the pressure  obtained from ruby at room temperature and the superconducting transition of a piece of lead revealed good agreement to within \SI{0.2}{\GPa} for the pressure cells used for magnetisation measurements.

The effect of different pressure media is discussed in S I of the supplementary information.

\subsection{Electrical Transport Measurements}
\label{subsec:Methods:Transport}
For the electrical measurements, six bilayer electrodes were deposited on one anvil in a three-step process without breaking vacuum. Firstly, the anvil was cleaned using an RF argon plasma etch, followed by sputtering \SI{20}{\nano\meter} of nichrome, and finally evaporation of \SI{150}{\nano\meter} gold. To ensure potential electrical shorts between electrodes, any nichrome overspray was removed using TFN etchant. 

Gold contacts were evaporated on top of the sample. Epo-Tek H20E silver paint was used to connect the samples to the electrodes on the anvil. 
A four-probe AC method was used to measure the resistance with a current $I = \SI{1}{\milli\ampere}$. The six electrodes were used to measure \Vl, the longitudinal and \Vt, the transverse voltages, respectively. The Hall coefficient was calculated from the antisymmetric part of $\Vt(H)$ under reversal of the magnetic field $H$ as 

\begin{equation*}
	\RH = \frac{\Vt(H)-\Vt(-H)}{2 H} \frac{t}{I} \quad .
\end{equation*}

For the electrical measurements, the gaskets were insulated using a mixture of Stycast epoxy 2850FT and BN powder; the mixture was pressed between the anvils to above the maximum pressure required for the experiments and then cured whilst pressurised. A \SI{400}{\micro\meter} hole through the insulation was drilled for the sample space.

\subsection{Magnetic Measurements}
\label{subsec:Methods:Magnetic}
A Quantum Design Magnetic Property Measurement System (MPMS) was used to measure the DC magnetic moment of the sample inside the pressure cell as detailed in section S IV of the supplementary information.
The transition temperature, \Tc, has been determined as the temperature where $\chi(T)$ has dropped by $\SI{10}{\percent}$ of the normalised step, i.e. close to the onset of the transition. This procedure results in uncertainty less than \SI{0.05}{\kelvin} of \Tc.

%
%
\section{Experimental Results}

\begin{figure}%
	\includegraphics[width=\figurewidth]{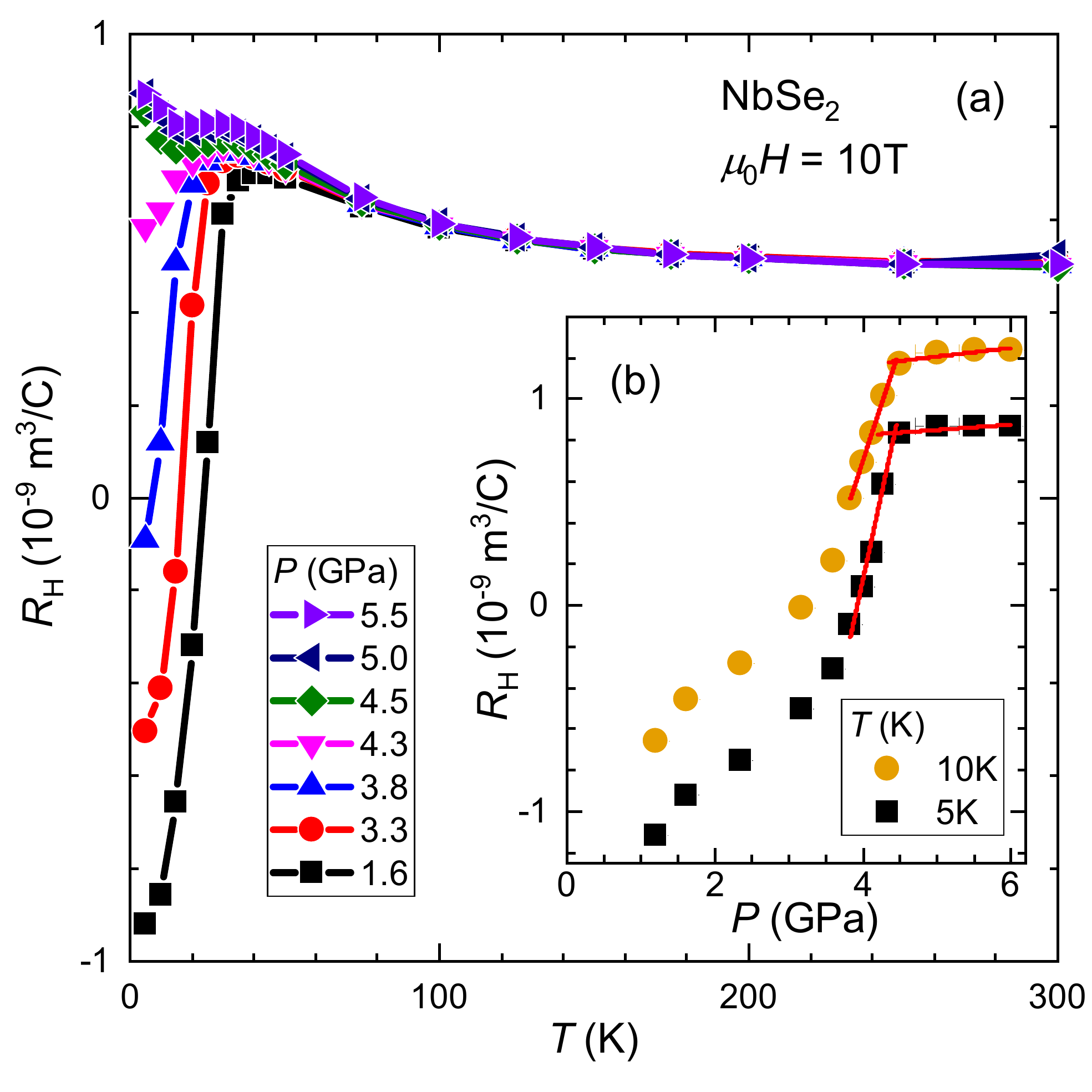}%
	\\
	\includegraphics[width=\figurewidth]{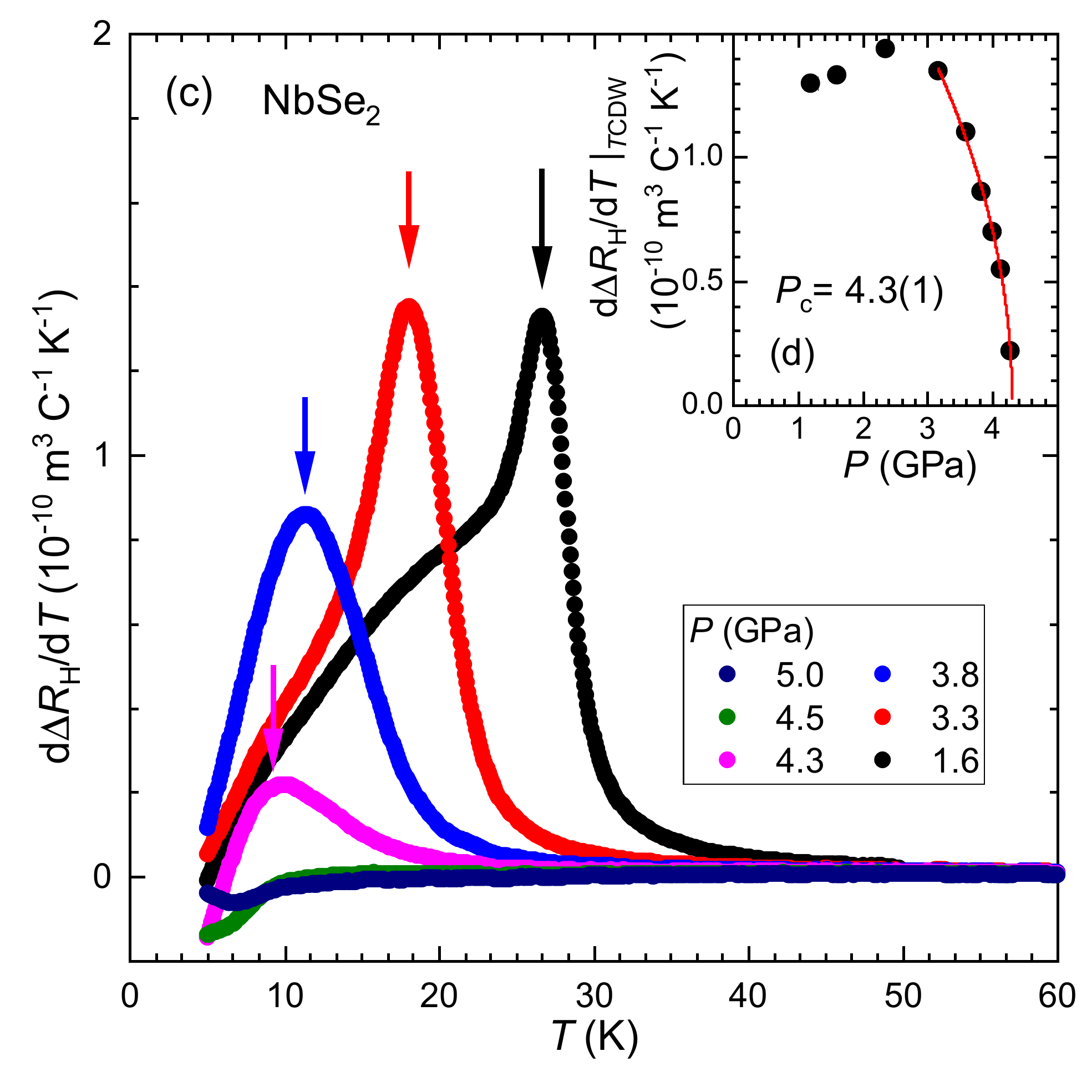}
	\caption{\textbf{Suppression of the CDW transition in \NbSe\ under pressure.} (a) The temperature dependence of the Hall coefficient for sample 1. (b) Pressure dependence  $\RH(P)$ measured at $\mu_0 H = \SI{10}{\tesla}$ at selected temperatures. Straight lines highlight linear fits to the data. (c) Derivative of the Hall coefficient $(\dd \Delta \RH / \dd T)$ was calculated after subtracting the high-pressure background, i.e. $\Delta \RH (T,P) = \RH(T,P) - \RH(T,\SI{5.5}{\GPa})$. Arrows indicate \TCDW\ extracted as the maximum. (d) Amplitude of the peak in $\dd \RH / \dd T$. Solid line denotes empirical power-law fit $\left.\dd\Delta\RH/\dd T\right|_{\TCDW} = A \left|P-\Pc\right|^b$ to the data above \SI{3}{\GPa} giving a critical pressure $\Pc=\SI{4.3(1)}{\GPa}$.}%
	\label{fig:RH}%
\end{figure}%

\begin{figure}%
	\includegraphics[width=\figurewidth]{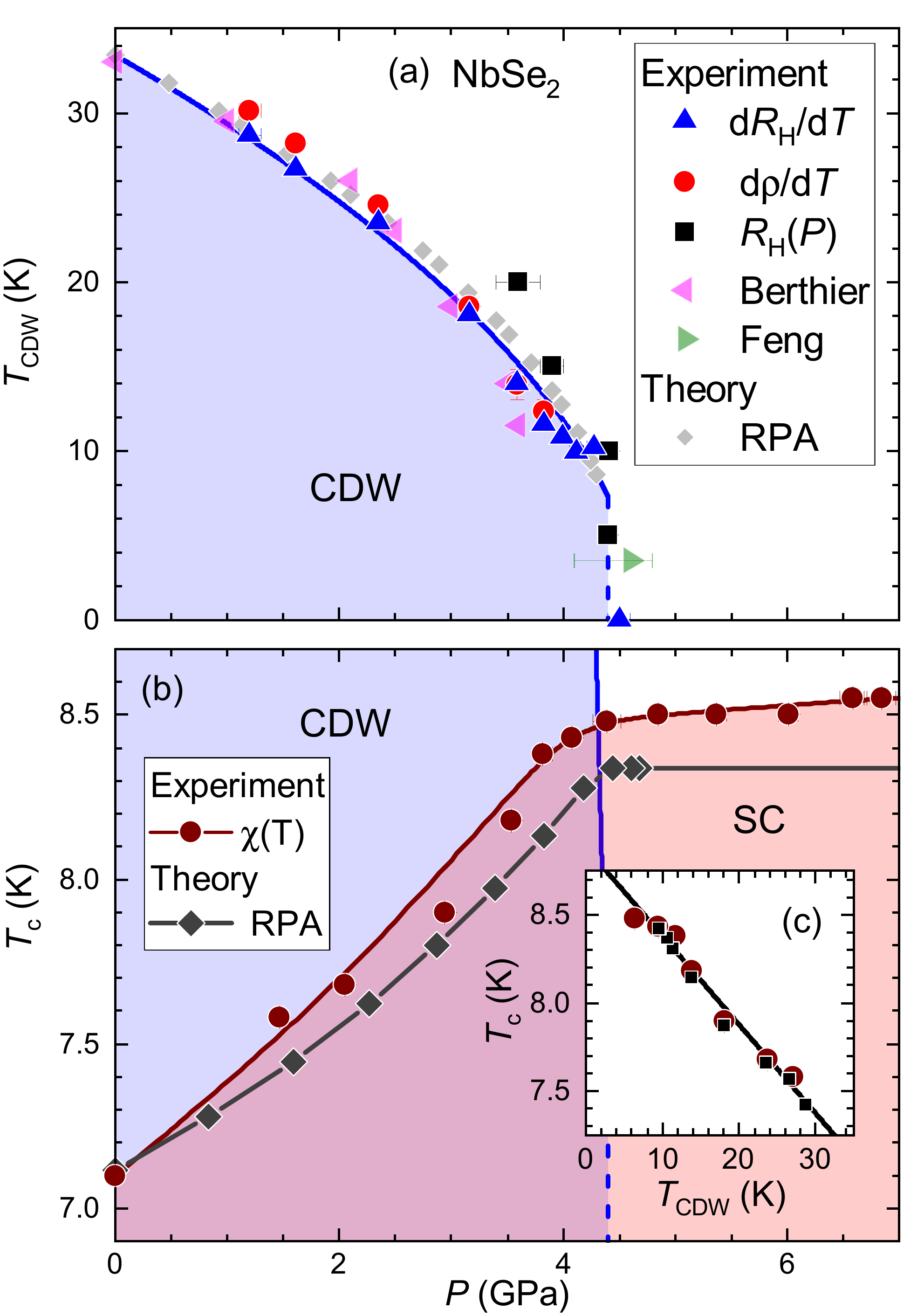}
	\caption{\textbf{High-pressure phase diagram of \NbSe.} 
	(a) 
	Experimental values of $\TCDW(P)$ are determined as the peak in $\dd \Delta\RH / \dd T$ as shown in \autoref{fig:RH}, as the minimum in $\dd \rho / \dd T$ as shown in SI, and the kink in $\RH(P)$ as shown the inset of \autoref{fig:RH}(a).
	Theoretical values of $\TCDW(P)$ are calculated as described in methods section S V and section S VI of the Supplementary Material.
	Data from Refs.~\cite{Berthier1976,Feng2012} are included. 
	The solid line marks a power-law fit to our experimental datasets of $\TCDW(P)$ for $P\leq\SI{4.3}{\GPa}$.
		The dashed line illustrates the sharp decrease of \TCDW\ at \SI{4.4}{\GPa} (see text). 
	(b)
	Experimental results for $\Tc(P)$ are extracted from magnetisation measurements (\autoref{fig:SC}(a)). A detailed comparison of $\Tc(P)$ using different pressure media is given in S I. 
	Theoretical values of $\Tc(P)$ are calculated as described in S VI.
	The boundary of the CDW phase is reproduced from (a).
	Inset (c) shows the relation of the superconductivity and CDW transition temperatures for two samples: Red circles and black squares denote \Tc\ of sample 1 and 2 detected with $\chi(T)$ and $\rho(T)$, respectively. The solid line is a linear fit.}%
	\label{fig:PD}
\end{figure}

Our high-pressure Hall effect measurements show the suppression of \TCDW\ under pressure in \autoref{fig:RH}(a). At high temperatures, the Hall coefficient, $\RH$, is weakly temperature dependent and does not change with pressure indicating that the electronic structure in the non-CDW state remains unchanged by pressure. At \TCDW, $\RH(T)$ shows a large drop and a sign change consistent with earlier results at ambient and low pressure \cite{Lee1970,Yamaya1972,Huntley1973}. Such a sign change has been linked to the CDW transition in a variety of systems including \NbSe\ \cite{Moncton1975,Dalrymple1984}, cuprate YBa$_2$Cu$_3$O$_y$ \cite{LeBoeuf2011}, and \TiSe\ \cite{Knowles2020} and has been confirmed by model calculations \cite{Evtushinsky2008}. 
The contribution of the CDW to the Hall coefficient $\Delta \RH(T,P) = \RH(T,P) -\RH(T,\SI{5.5}{\GPa})$, is calculated by subtracting the non-CDW form well above the critical pressure. In the derivative, $\dd \Delta\RH / \dd T$, the CDW transition manifests as a pronounced peak as shown in \autoref{fig:RH}(c). $\TCDW(P)$ associated with the maximum in $\dd \Delta\RH / \dd T$ shifts to lower temperature as pressure is increased in good agreement with $\TCDW(P)$ extracted from resistivity measurements as well as with previous results of \TCDW\ as highlighted in \autoref{fig:PD}(a). The benefit of analysing the Hall coefficient is that the strong signature can be traced to higher pressures where the signature in resistivity is lost \cite{Berthier1976}. We observe the CDW transition in $\Delta \RH(T)$ up to a pressure of \SI{4.3}{\GPa}.

The CDW transition temperature drops abruptly above \SI{4.3(1)}{\GPa} as can be seen  from both the isobaric temperature dependence and the isothermal pressure dependence of the Hall coefficient.
The peak in the isobaric temperature dependence $\dd \Delta\RH / \dd T$ is reduced in amplitude above \SI{3}{\GPa} as shown in \autoref{fig:RH}(c). In fact, the reduction is most consistent with a power-law suppression where the amplitude vanishes at \SI{4.3(1)}{\GPa} suggesting an absence of the CDW above this pressure. Thus, we conclude that the CDW signature is absent from the temperature dependence of the Hall effect and highlight this in the phase diagram  as $\TCDW(\SI{4.4}{\GPa})=0$ (blue triangle at \SI{4.4}{\GPa} in \autoref{fig:PD}(a)).
The isothermal pressure dependence $\RH(P)$ exhibits a pronounced kink associated with the critical pressure of the CDW phase, $\PCDW(T)$ (cf.\ intersecting linear fits in \autoref{fig:RH}(b)). The position of $\PCDW(T)$ is included as black squares in the phase diagram in \autoref{fig:PD}. $\PCDW(T)$ becomes independent of temperature for $T\leq\SI{10}{\kelvin}$, i.e. the kink in $\RH(P)$ is found at the same pressure $\PCDW(T)=\SI{4.4}{\GPa}$ for \SI{5}{\kelvin} and \SI{10}{\kelvin}. In section SI of the supplementary information we show that this result is also true if the Hall effect is probed in smaller magnetic fields.

\begin{figure}%
	\includegraphics[width=\figurewidth]{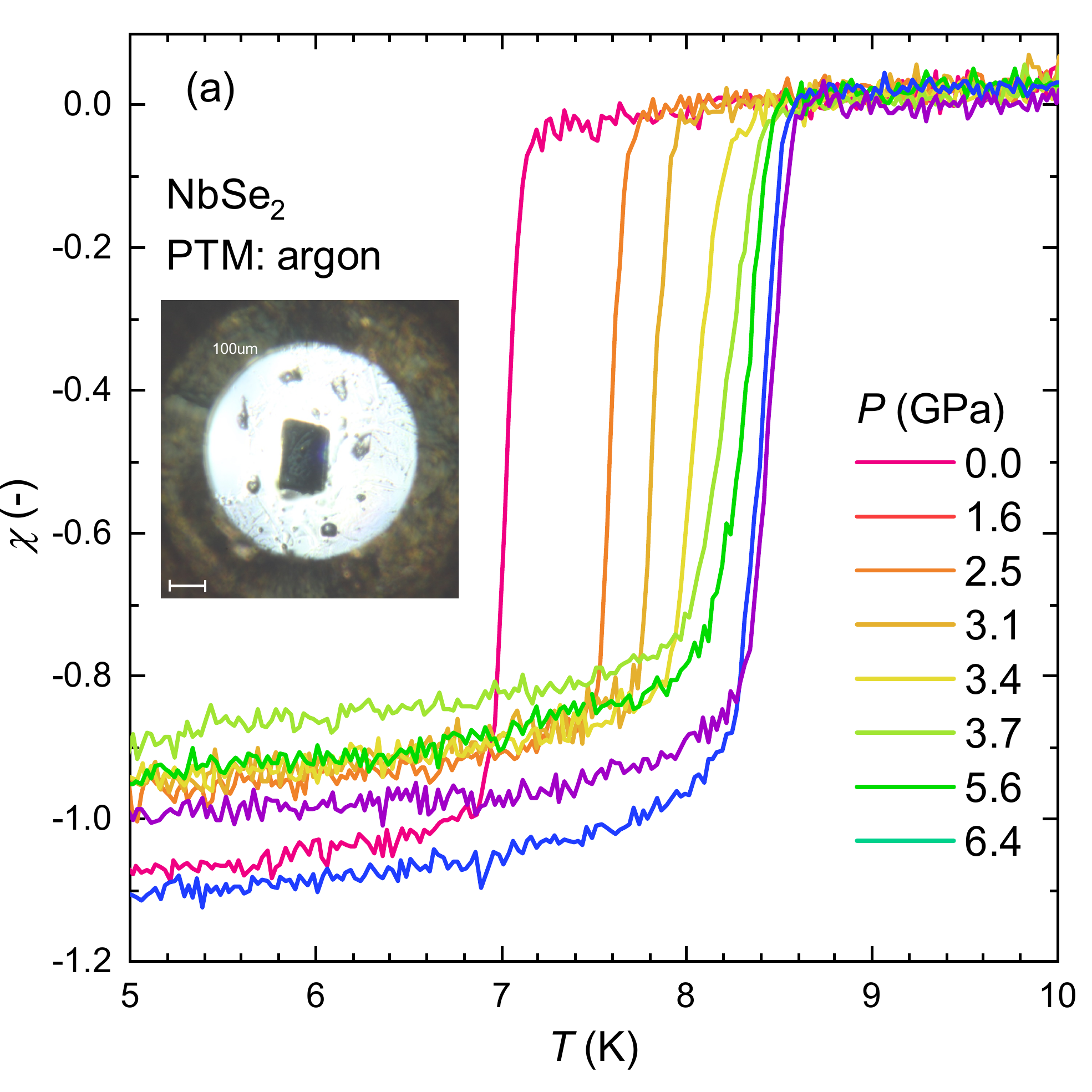}%
	\\
	\includegraphics[width=\figurewidth]{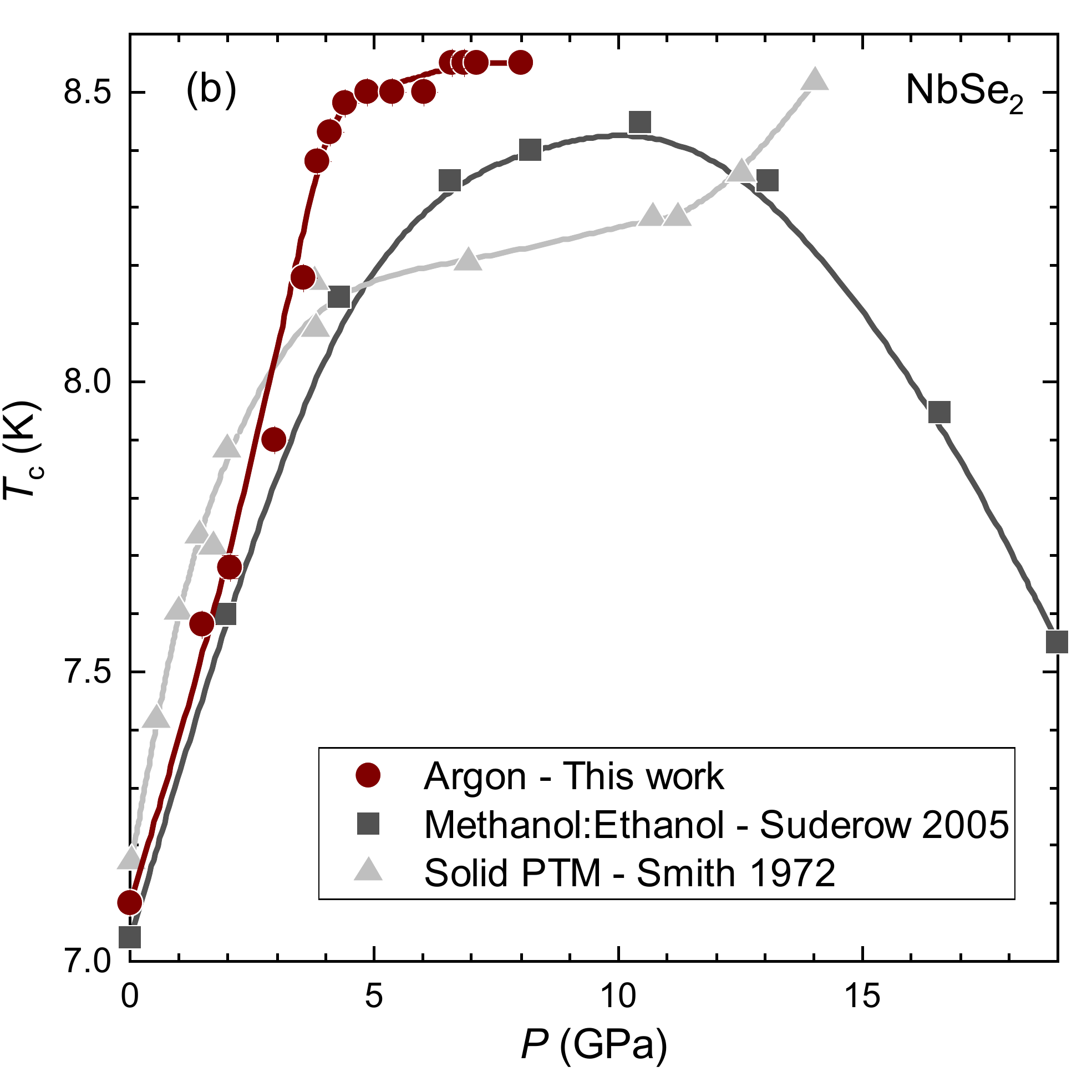}%
	\caption{\textbf{Enhanced superconductivity in \NbSe\ under pressure.} (a) The volume susceptibility $\chi$ measured on warming for sample 2 in a magnetic field $\mu_0 H = \SI{0.5}{\milli\tesla}$ after zero-field cooling. The transition into the superconducting state inferred from the diamagnetic signal shifts to higher temperatures as the pressure is increased. Inset shows a picture of the sample and ruby chips inside the pressure cell. Argon was used as a pressure medium. (b) Comparison of our measurements with previous studies beyond the CDW critical pressure \cite{Smith1972,Suderow2005}.}%
	\label{fig:SC}%
\end{figure}

Superconductivity is boosted under pressure in clear anticorrelation to  the CDW. We trace $\Tc(P)$ as the onset of the diamagnetic signal in magnetic susceptibility measurements, $\chi(T)$, as presented in \autoref{fig:SC}(a). The sharp onset gives $\Tc=\SI{7.1}{\kelvin}$ at ambient pressure  in good agreement with our resistivity measurements (cf.~S II of the supplementary information) and other published work e.g. Refs. \onlinecite{Wilson1969,Wieteska2019,Rohr2019}. With increasing pressure, $\Tc(P)$ shifts to higher temperature whilst the transition remains very sharp. Above \SI{4.4}{\GPa}, $\Tc(P)$ saturates at \SI{8.5}{\kelvin}. The measurements presented in \autoref{fig:SC}(a) used argon as a pressure transmitting medium (PTM) which remains hydrostatic up to \SI{11}{\GPa} \cite{Tateiwa2009}. We find very good agreement with $\Tc(P)$ extracted from our resistivity measurements up to \SI{5.5}{\GPa} -- the limiting pressure for hydrostaticity of the PTM glycerol used for the electrical transport measurements \cite{Tateiwa2009}. 

Our measurements of $\Tc(P)$ differ significantly from the two previous studies which extend beyond the critical pressure of the CDW by Smith et al.\ \cite{Smith1972} and Suderow et al.\ \cite{Suderow2005} as presented in \autoref{fig:SC}(b). A similar initial slope of $\Tc(P)$ is seen in our work and those previous studies (including further work limited to below \SI{3}{\GPa} \cite{Rohr2019,Berthier1976}). However, a marked difference is observed above \SI{4}{\GPa} where all datasets show a plateau at different values of \Tc. Notably, we observe the highest \Tc\ in any of the measurements. Given that impurities have been shown to cause a suppression of superconductivity outside the CDW phase \cite{Cho2018,Dalrymple1984}, this suggests that the studies by Smith et al and Suderow et al. suffered from sample impurities or inhomogeneities. Our electrical resistivity measurements reveal that the residual resistance ratio remains large at above 60 while Smith et al. and Suderow et al. have not provided a characterisation of their samples at ambient pressure  and could not monitor the pressure inhomogeneity effects with the ac susceptibility measurements. In fact, Smith et al. used non-hydrostatic solid pressure medium which is know to lead to pressure inhomogeneities and anisotropy. In the case of Suderow et al. methanol:ethanol was used with a hydrostatic limit of $\approx\SI{9}{\GPa}$ \cite{Tateiwa2009}. However, we show in the supplementary information SII that the sample preparation and stresses from the sample touching the gasket can lead to a reduced \Tc\ outside the CDW phase and we reproduce the $\Tc(P)$ of Suderow in a sample with a broadened transition. In summary, we argue that our data for the first time reveal the intrinsic high-pressure evolution of the superconducting transition temperature of \NbSe.

\begin{figure}%
	\includegraphics[width=\figurewidth]{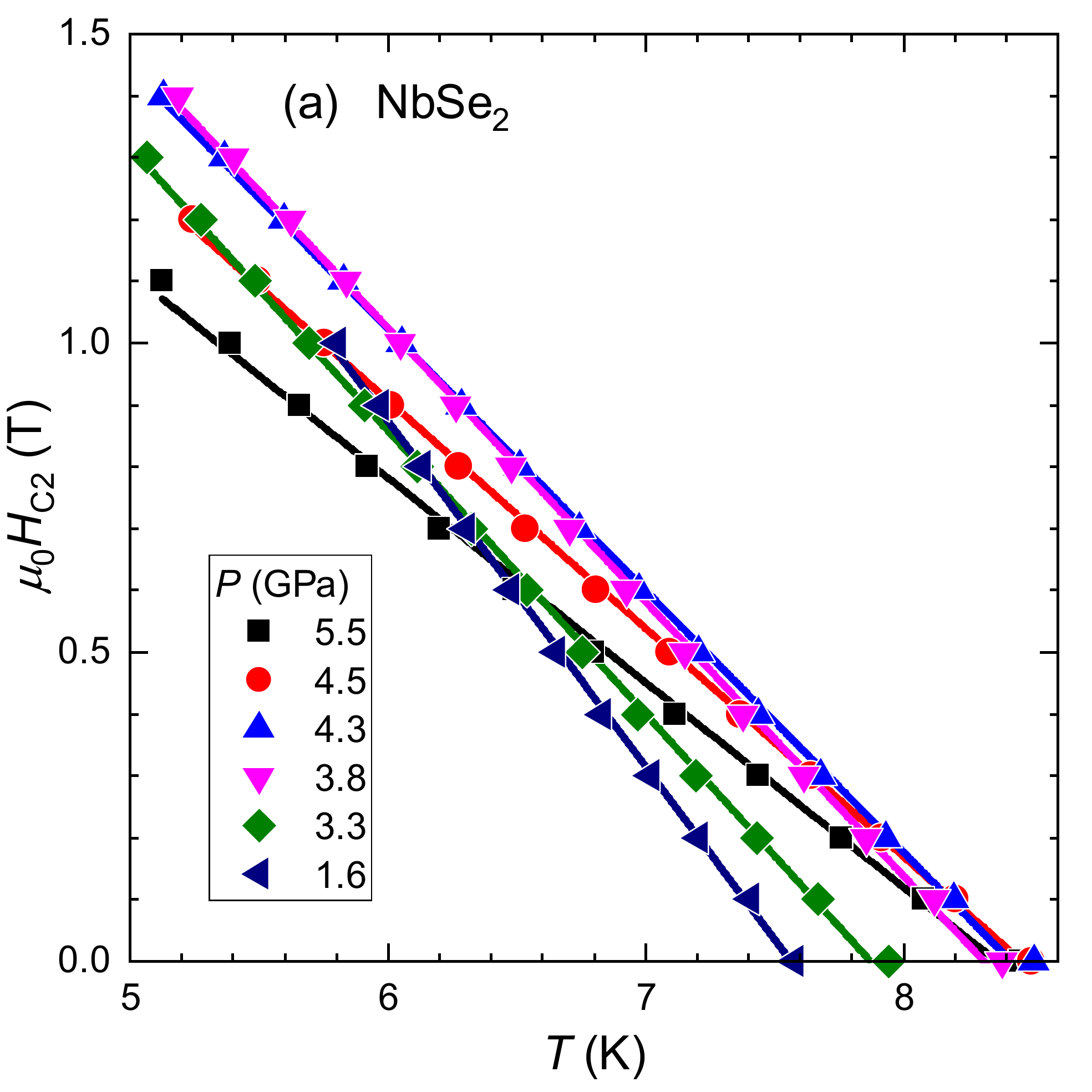}
	\\
	\includegraphics[width=\figurewidth]{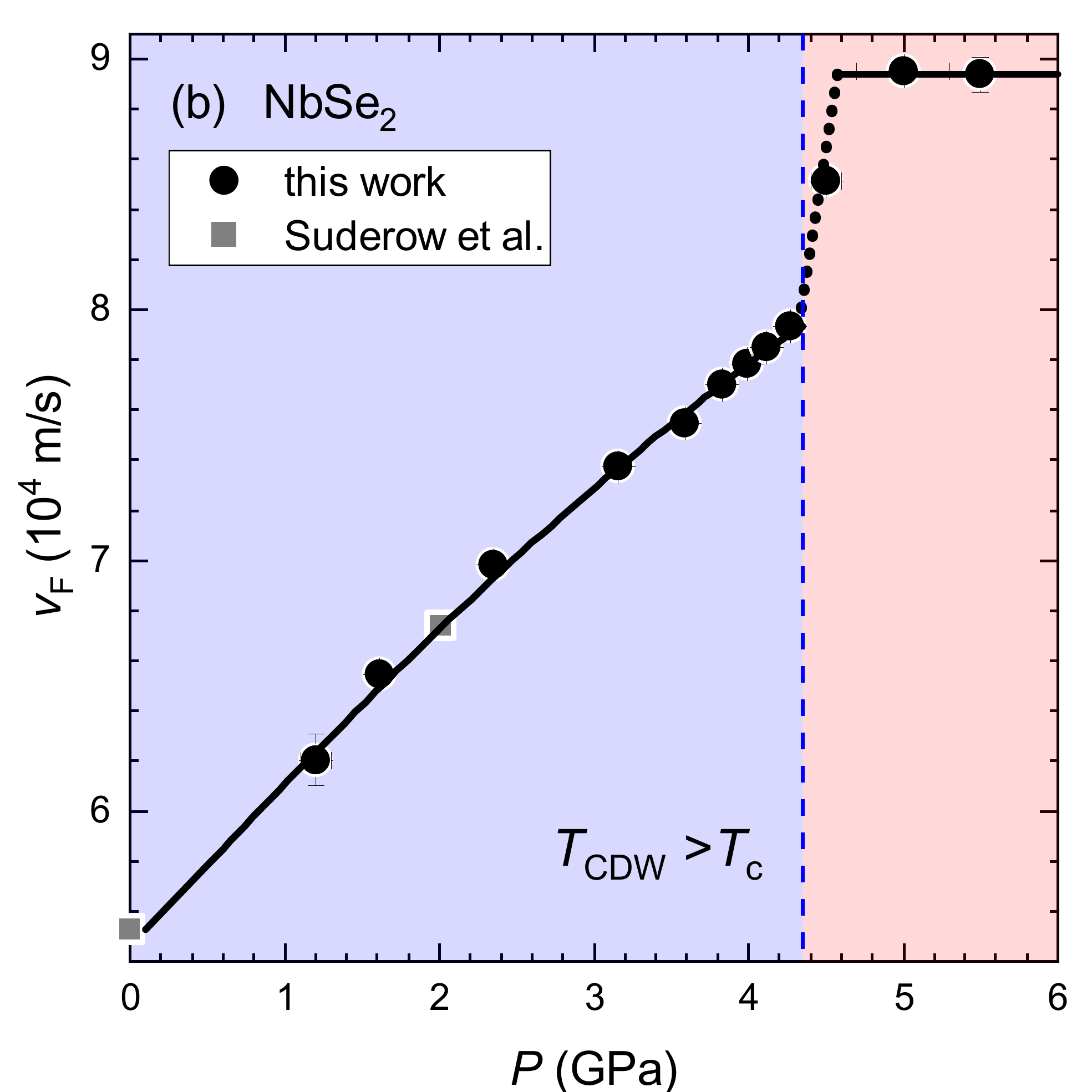}
\caption{\textbf{Upper critical field and Fermi velocity in \NbSe.} 
(a)  Linear fits to $\Hcii(T)$ have been used to extract $\vF(P)$ shown in (b). For details see S III of the supplementary information.  Data of Suderow et al. \cite{Suderow2005} have been included in (b) for the pressure range with good agreement of \Tc\ (see SII of the supplementary information). Lines are guides to the eye.
}%
\label{fig:Hc}%
\end{figure}%

We use measurements of the upper critical field \Hcii\ to characterise the Fermi velocity, \vF, on the strong coupling Nb cylinders. 
At all pressures, we observe a linear dependency of $\Hcii(T)$ below \Tc\ as presented in \autoref{fig:Hc}(a) and in agreement with earlier ambient-pressure experiments \cite{Toyota1976,Dalrymple1984}. We use the slope $\left.\dd \Hcii / \dd T \right|_{\Tc}$ to calculate the Fermi velocity, \vF, shown in  \autoref{fig:Hc}(b). Both \textit{ab initio} calculations and analysis of the Usadel equations show that $\left.\dd \Hcii / \dd T \right|_{\Tc}$ for $H$ along the crystallographic $c$ direction, used here, is almost entirely determined by the strong coupling Nb cylinders \cite{Arai2004,Golubov2003}.
In addition, we find perfect agreement with \vF\ associated with the Nb cylinders identified in the previous high-pressure measurements by Suderow et al. at low pressures in the pressure  range where \Tc\ of Suderow's sample is in agreement with our measurements \cite{Suderow2005} (grey squares in \autoref{fig:Hc} (b)). 

At the critical pressure of the CDW phase, we observe a jump in the Fermi velocity suggesting a collapse of the CDW gap (dotted line in \autoref{fig:Hc}(b)). Initially, a steady increase of \vF\ by \SI{40}{\percent} is observed for $P\leq \SI{4.3}{\GPa}$, i.e. where $\Tc < \TCDW$. This initial steady rise is most naturally associated with  a continuous shrinkage of the CDW gap and reduction of the average re-normalisation on the niobium bands.  By contrast, a jump of $\approx \SI{10}{\percent}$ is observed over a narrow pressure range at $P\sim\SI{4.4}{\GPa}$, exactly at the pressure where \TCDW\ drops to zero consistent with a collapse of the CDW gap.  Above \SI{5}{\GPa} \vF\ saturates suggesting that the coupling of the CDW mode to the electronic states quickly reduces outside the CDW phase. 
As $\Hcii(T)$ is well fitted by a linear dependence to lowest magnetic fields (i.e. right to \Tc), we conclude that the CDW collapse is also present in zero magnetic field at \SI{4.4}{\GPa} corroborating the evidence for a drop in \TCDW\ from the Hall-effect measurements in magnetic fields above \SI{2}{\tesla} discussed above.

%
%
\section{Theoretical Modelling}
We find that the stiffening of the bare longitudinal acoustic phonon from which the CDW develops can account for the suppression of \TCDW\ under pressure. We use electron-phonon coupling dependent on the ingoing and outgoing momentum and the specific shape of the Fermi surface of \NbSe\ including the orbital character. The model was developed earlier by one of us as outlined in sections S V  and S VI of the supplementary material  \cite{Flicker2015,Flicker2016}. 
In our RPA calculations, the overall magnitude of the electron-phonon coupling $\vec{g}$ is constrained to reproduce $\TCDW(P=0)=\SI{33.4}{\kelvin}$ (cf.~S III of the supplementary information) and we keep $\vec{g}$ fixed for all pressures.
To describe $\TCDW(P)$, we assume a linear stiffening of the longitudinal acoustic phonons underlying the CDW formation consistent with high-pressure inelastic x-ray studies \cite{Leroux2015} as detailed in section S V of the supplemental information. 
In \autoref{fig:PD}, the experimental transition temperatures are compared to model calculations.
From the good match with the experimental phase boundary up to \SI{4.3}{\GPa}, we conclude that the suppression of the CDW is indeed driven by the increase of the bare phonon frequency whilst the electron-phonon constant remains  unchanged.
While our model tracks the phase boundary well for $P<\SI{4.3}{\GPa}$, it is too simple to account for a possible change in the order of the transition.
Features omitted from the model which could account for such a change  include higher-order lattice coupling, fluctuation effects, or the effect of pressure on the electronic bandstructure.

\begin{figure}%
\includegraphics[width=\figurewidth]{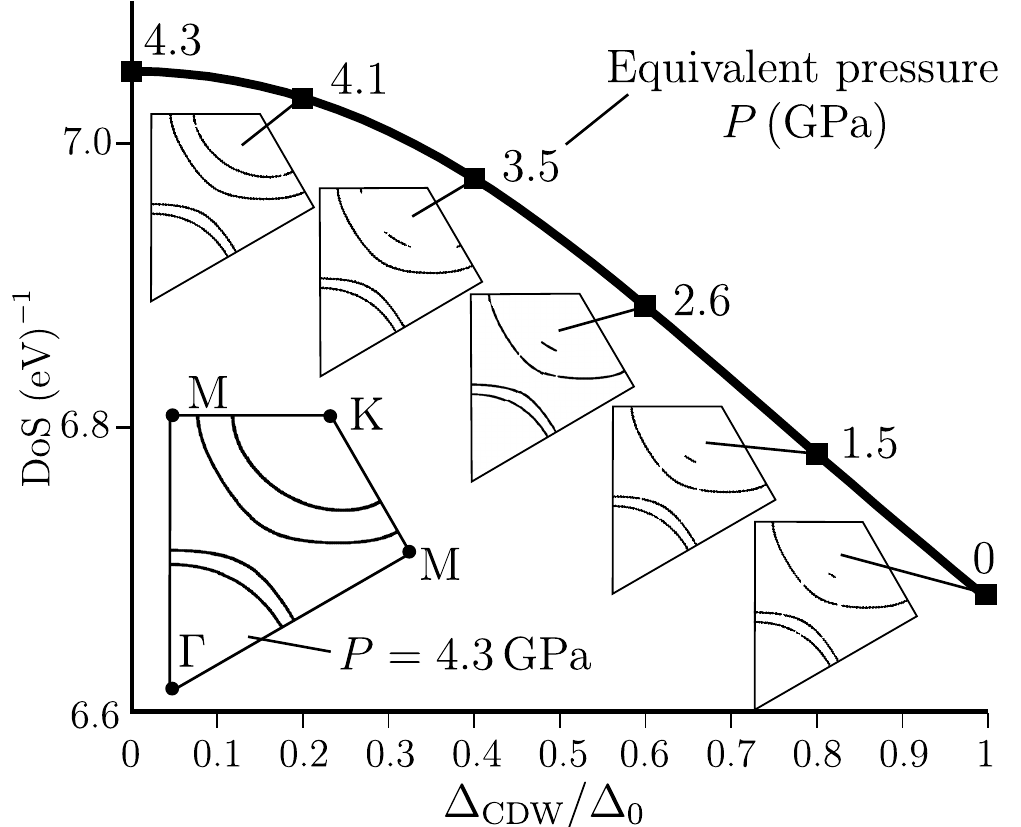}%
\caption{Reduction of DOS as a function of CDW gap magnitude for the two Nb-derived bands at \EF. Insets show the Fermi surface in a wedge of the Brillouin zone for specific values of \DeltaCDW\ and equivalent pressure. The plotted points were identified as the the points at which the RPA spectral function is within \SI{15}{\percent} of its maximum value.}%
\label{fig:FS}%
\end{figure}

A partial competition for density of states (DOS) is the main driver for the evolution of $\Tc(P)$. We use the experimentally determined phase boundary (solid line in ~\autoref{fig:PD}(a)) to scale the evolution of the CDW phase to our pressure data as detailed in section S VI of the supplemental information.
Inside the CDW phase, the DOS available for superconductivity is reduced due to the gapping of the inner K-pockets of the Fermi surface as illustrated in \autoref{fig:FS} leading to a reduction of \Tc.  As the CDW gap becomes smaller, the DOS available for superconductivity becomes larger which in turn accounts for almost the entire increase of \Tc\ and naturally explains why \Tc\ saturates above \PCDW\ as can be seen in \autoref{fig:PD}(b). Thus, we conclude that it is a  competition for DOS which suppresses \Tc\ inside the CDW phase.

%
%
\section{Discussion and Conclusions}

Our study provides the most comprehensive and consistent dataset of the boundary of the CDW phase in \NbSe\ to date and suggests a \nth{1}-order transition at $\PCDW=\SI{4.4}{\GPa}$.
The combined evidence of a drop in $\TCDW(P)$ at \PCDW\ extracted from the isobaric and isothermal analysis of the Hall effect and  the jump in $\vF(P)$ at \PCDW\ suggest that the  \nth{1}-order transition at \PCDW\ is an intrinsic characteristic of the CDW  in \NbSe.

Our work is the first high-pressure study to find indications of a \nth{1}-order transition at the critical pressure of the CDW order. The previous pressure study by Berthier et al. traced \TCDW\ to \SI{3.6}{\GPa} \cite{Berthier1976}. Only one further study has obtained information about the boundary of the CDW phase above \SI{3.7}{\GPa}: the X-ray measurements of Feng et al. \cite{Feng2012,Feng2015} who have interpreted a kink in the pressure dependence of the $c$-axis lattice constant at \SI{4.6}{\GPa} and \SI{3.5}{\kelvin} as the critical pressure of the CDW phase (green triangle in \autoref{fig:PD}(a)). An uncertainty has not been given by Feng et al., however, taking into account the scatter of $c(P)$ we estimate an uncertainty of 4.6(-0.5)(+0.2)\si{\GPa} as included in our phase diagram \autoref{fig:PD}(a). Taking into account this uncertainty, the data of Feng et al. are consistent with the critical pressure of \SI{4.4}{\GPa} found in our study.

The XRD measurements of Feng et al.\ cannot rule out a \nth{1}-order transition at \SI{4.4}{\GPa} \cite{Feng2012,Feng2015}. The in-plane lattice constant $a$ was measured with very high precision up to \SI{8.5}{\GPa} but does not reveal any signature at \PCDW\ and thus cannot be used to discriminate the order of the quantum phase transition. Together the lattice constants give an upper limit of \SI{0.5}{\percent} for a discontinuity in the volume leaving the possibility of a weak \nth{1}-order transition compatible with the small $\TCDW \approx \SI{10}{\kelvin}$ just below \PCDW.
Scaling in the X-ray diffraction measurements of Feng et al. is cut off above \SI{4}{\GPa} \cite{Feng2012,Feng2015}. At \SI{4.55}{\GPa} the XRD measurements find a cut-off of the  divergence of the inverse static correlation length and an order-of magnitude drop of the CDW intensity.  Only the low-pressure XRD data show clear evidence of \nth{2}-order behaviour (divergence of the inverse static correlation length and smooth decrease of the intensity of the CDW reflections up to \SI{4}{\GPa}) consistent with our continuous suppression of $\TCDW(P)$ up to \SI{4.3}{\GPa}.

Indications of a \nth{1}-order transition were observed before in electron-irradiated \NbSe\ but were attributed to disorder effects \cite{Cho2018}. 
Our samples preserve the high residual resistance ratio across \PCDW\ thus disorder is of negligible effect in our study. Instead, our results suggest that quantum fluctuations or coupling to the lattice may induce a \nth{1}-order transition in \NbSe.
Indeed, a suppression of  CDW order by quantum fluctuations was  proposed for \NbSe\ \cite{Feng2015,Leroux2015} and a strong coupling to the lattice plays a crucial part for the formation of the CDW in \NbSe\ \cite{Weber2011a}.
A weak \nth{1}-order phase transition to the incommensurate CDW state was observed in related CDW systems \TiSe\
and 1T-TaS$_2$\cite{Joe2014,Moncton1975}. Thus, our results suggest that quantum fluctuations and/or coupling to the lattice play an important role  when suppressing CDW order to zero temperature and induce a \nth{1}-order transition  in many CDW materials. 

Our detailed measurements lead us to different conclusions on the interplay between superconductivity and CDW order compared to previous studies. The increase in \Tc\ is in clear anticorrelation with $\TCDW(P)$ as highlighted in \autoref{fig:PD}(c). In addition, $\Tc(P)$ is virtually constant for $P\geq\SI{4.4}{\GPa}$, i.e. outside the CDW phase. This anticorrelation and the saturation are clear signs that the superconductivity is in competition with the CDW phase as suggested for CDW superconductors in general and \NbSe\ in particular in previous studies \cite{Bilbro1976,Berthier1976}. Our model calculations show quantitatively that the suppression is caused by the depletion of density of states inside the CDW phase. Notably, we do not observe a maximum in \Tc\ around the critical pressure of the CDW phase. Thus, we can rule out a boost to \Tc\ from quantum critical fluctuations. Similarly, we can rule out a boost to superconductivity from the presence of the static CDW as suggested by Kiss et al. \cite{Kiss2007}: Such a boost should manifest in a correlation of \TCDW\ and \Tc\ and a drop of \Tc\ at \PCDW\ neither of which  is observed. Furthermore, our data suggest  that a notable maximum is absent for pressures below \SI{10}{\GPa} thus suggesting that the electron-phonon coupling responsible for the superconductivity is largely independent of pressure. 

The \nth{1}-order nature of the CDW close to \PCDW\ maybe the reason for the absence of a dome-shaped enhancement in $\Tc(P)$ upon suppression of the CDW. 
At a \nth{2}-order quantum phase transition, i.e. a quantum critical point, a superconducting dome was observed in many systems including at the CDW QCP in \TiSe \cite{Morosan2006a}
, close to the structural QCP in (Sr,Ca)$_3$Ir$_4$Sn$_{13}$ \cite{Klintberg2012}, and at the antiferromagnetic QCP in CePd$_2$Si$_2$ \cite{Mathur1998}. 
By contrast, dome-shaped superconducting phases are usually absent in systems with \nth{1}-order quantum phase transitions as shown for instance in UGe$_2$ \cite{Saxena2000}. This is a clear indication that the gapped fluctuation spectrum at \nth{1}-order quantum phase transitions is not suitable to mediate or enhance superconductivity. Our data suggest that this principle applies to \NbSe, too.

While we have shown clearly that CDW is suppressing superconductivity, our data also reveal that superconductivity has no effect on \TCDW.
Firstly, the gradual suppression of \TCDW\ for $P<\SI{4}{\GPa}$ cannot be driven by the superconductivity because $\TCDW>\Tc$.  Secondly, whilst the drop of \TCDW\ occurs at the pressure where the power-law fit predicts $\TCDW<\Tc(H=0)$, our Hall-effect data have detected the CDW transition in high magnetic fields where  superconductivity is suppressed. 
In addition, the suppression of the amplitude (peak height) of the CDW signature in $\RH(T)$ is based entirely on data where $\TCDW>\Tc(H=0)$. 
Thus, our data suggest that the drop of \TCDW\ at \SI{4.4}{\GPa} is intrinsic to the CDW in \NbSe\ and not driven by competition with superconductivity. 
Hence, we conclude that the competition between superconductivity and CDW is mostly unidirectional in \NbSe\ with only superconductivity suppressed by CDW but not the other way around.
Such a unidirectional competition is supported by previous X-ray measurements at ambient pressure which show that the intensity of the CDW reflection is not reduced at \Tc\ in zero field and not enhanced upon suppressing superconductivity in large magnetic fields \cite{Du2000,Moncton1975}. 

In summary, our results lead to several profound conclusions about the interplay of CDW order and superconductivity in \NbSe. 
(i)  Superconductivity is suppressed inside the CDW phase due the reduced electronic density of states available for superconductivity,
(ii) Superconductivity is not reducing \TCDW. Instead the suppression of \TCDW\ under pressure is consistent with the stiffening of the underlying bare phonon mode.
(iii) \TCDW\ drops abruptly at $\PCDW = \SI{4.4}{\GPa}$ indicating a \nth{1}-order transition. 
(iv) Superconductivity is not enhanced at \PCDW\ potentially due to CDW fluctuations being cut off at the \nth{1}-order transition.

%
%
\begin{acknowledgments}
The authors would like to thank Jasper van Wezel, Jans Henke, Nigel Hussey, Hermann Suderow, and  Antony Carrington for valuable discussion. The authors acknowledge supported by the EPSRC under grants EP/R011141/1, EP/L025736/1, EP/N026691/1 as well as the ERC Horizon 2020 programme under grant 715262-HPSuper.
\end{acknowledgments}

\section*{Additional information}
Data are available at the University of Bristol data repository, \href{https://data.bris.ac.uk/data/}{data.bris}, at \url{https://doi.org/10.5523/bris.xxxx} .

%
%
\bibliography{Moulding_NbSe2}



\section*{Supplementary material}

# Supplementary Information for
# "Absence of superconducting dome at the charge-density-wave quantum phase transition in 2H-NbSe₂"

Owen Moulding,[1] Israel Osmond,[1] Felix Flicker,[2] Takaki Muramatsu,[1] and Sven Friedemann[1, *]

[1] HH Wills Laboratory, University of Bristol, Bristol, BS8 1TL, UK

[2] Rudolf Peierls Centre for Theoretical Physics, University of Oxford, Department of Physics,
Clarendon Laboratory, Parks Road, Oxford, OX1 3PU, United Kingdom

(Dated: Wednesday 11th November, 2020)

## S I. HIGH-PRESSURE HALL EFFECT MEASUREMENTS

The phase boundary $T_{CDW}(P)$ has been extracted from both $d\Delta R_H/dT$ and $R_H(P)$ presented for measurements in $\mu_0 H = 10\,T$ in the main manuscript. An important observation is the abrupt drop of $T_{CDW}$ at 4.4 GPa. This is best seen in $R_H(P)$ in the inset of Fig. 1 of the main text.

In Fig. S1 we show $R_H(P)$ extracted at $\mu_0 H = 2\,T$, i.e. a field just above the critical field of the superconducting state. We find the same behaviour: A kink in $R_H(P)$ at 4.4 GPa which does not shift between 5 K and 10 K, i.e. a very steep phase boundary of $T_{CDW}(P)$ at 4.4 GPa. Thus, we observe that the drop in $T_{CDW}(P)$ is not due to the effect of a finite magnetic field on the CDW transition. This argument is extended to lower fields by the observation of linear $H_{c2}(T)$ at 4.5 GPa presented in Fig. 4(a) with no indications of a change in slope expected for a transition from the CDW state to the pure superconducting state (cf. Fig. 4(b) of the main text).

In Fig. S2 we show $R_H(P)$ for further temperatures. At $T \leq 10\,K$, a single kink is observed in $R_H(P)$. At 15 K and 20 K, two kinks are visible in $R_H(P)$. The kink at 3.8(1) GPa and 3.5(2) GPa, respectively from the steepest slope to a very much reduced slope matches with the phase boundary $P_{CDW}(T)$. At $T \leq 10\,K$ the steepest part of $R_H(P)$ has a similar steep slope compared to 15 K and 20 K. Thus, we identify the low-pressure kink at 15 K and 20 K and the single kink at $T \leq 10\,K$ with the boundary of the CDW phase.

At 15 K and 20 K, a second kink at 4.4 GPa towards constant $R_H(P)$ happens at the same pressure like the single kink at 5 K and 10 K. The origin of this second kink remains elusive and we can only speculate that it is related to fluctuations of the CDW order for pressures below 4.4 GPa. In fact, strong fluctuations have been inferred from ARPES experiments [1] and mode-mode-coupling calculations [2].

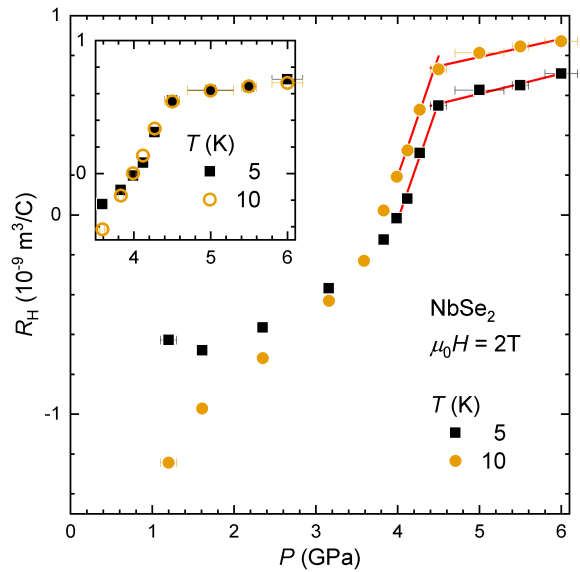

FIG. S1. Pressure dependence of the Hall coefficient at $\mu_0 H = 2\,T$ for $T = 5\,K$ and $T = 10\,K$. Straight lines show linear fits. The T=10 K data has been offset in the inset to match the high-pressure value of the T=5 K data.

## S II. INFLUENCE OF PRESSURE MEDIA AND SAMPLE PREPARATION ON SUPERCONDUCTIVITY

We have studied the influence of different pressure media and sample preparation on superconductivity under pressure in 2H-NbSe₂. In Fig. S3(a) we provide a comparison of two different pressure media used in this study: argon and glycerol. Argon was used for the measurements of the magnetic susceptibility presented in Fig. 3 of the main manuscript. Glycerol was used for all electrical transport measurements, e.g. Fig. 1 of the main manuscript. A second measurement of the magnetic susceptibility was done with glycerol as a pressure medium as presented in Fig. S4.

The comparison of $T_c(P)$ from these three measurements shows very good agreement up to $P = 5.5\,GPa$ (cf. Fig. S3(a)). This corresponds to the hydrostatic limit of glycerol which undergoes a glass transition around this pressure at 300 K [3]. Above 5.5 GPa, $T_c(P)$ is reduced for the samples in glycerol pressure medium. We at-

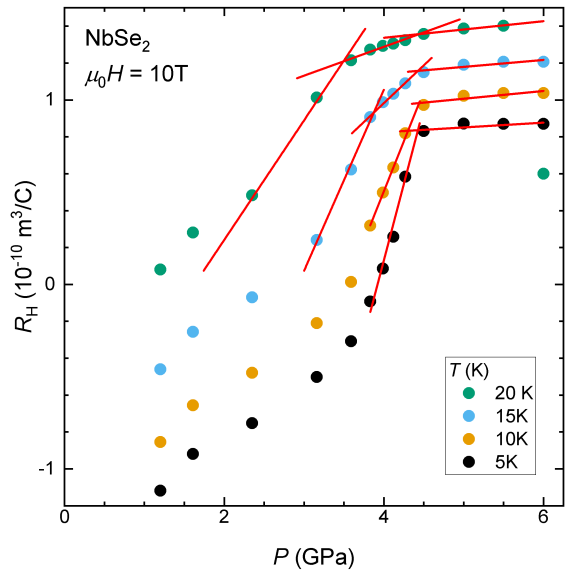

FIG. S2. Pressure dependent Hall coefficient $R_H(P)$ at selected temperatures. Straight lines highlight linear fits to the data.

tribute this reduction of $T_c$ to a uniaxial compression along the crystallographic $c$ axis of the sample as samples were mounted with the crystallographic $c$-direction perpendicular to the anvil cutlets [4, 5]. Pressure has been applied at room temperature and the pressure cells have been warmed to $300(1)\,\mathrm{K}$ after application of pressure. We conclude that glycerol provides high-qualtiy hydrostatic conditions for 2H-NbSe$_2$ up to 5.5 GPa.

Our measurements of $T_c(P)$ differ significantly from previous studies extending to beyond the critical pressure of the CDW as highlighted in Fig. S3(b) [6, 7]. In order to identify the cause for these differences, we have studied the relevance of the pressure medium and sample preparation. As discussed above, we find a reduced $T_c$ in non-hydrostatic conditions for glycerol above 5.5 GPa. Thus, we conclude that the reduced $T_c$ observed by Smith et al [6] is due to the usage of a solid pressure medium.

Suderow et al. used methanol:ethanol which provides good hydrostatic conditions up to 10 GPa [8]. Yet, $T_c$ is reduced and a much smoother rise of $T_c(P)$ is observed by Suderow et al.. We could reproduce the behaviour seen by Suderow et al. in one measurement using pentane:isopentane as a pressure medium (sample 1 in Fig. S3(b)). Pentane:isopentane is very similar to methanol:ethanol and provides good hydrostatic conditions to 10 GPa [8]. A second sample measured in pentane:isopentane, however, followed the $T_c(P)$ of our argon measurements (sample 2 in Fig. S4(b)). Thus, we conclude that the differences in $T_c(P)$ are not due to the pressure medium used as long as it provides good hydrostatic conditions.

We could identify a difference in the sharpness of the superconducting transition in $\chi(T)$ to correlate with the

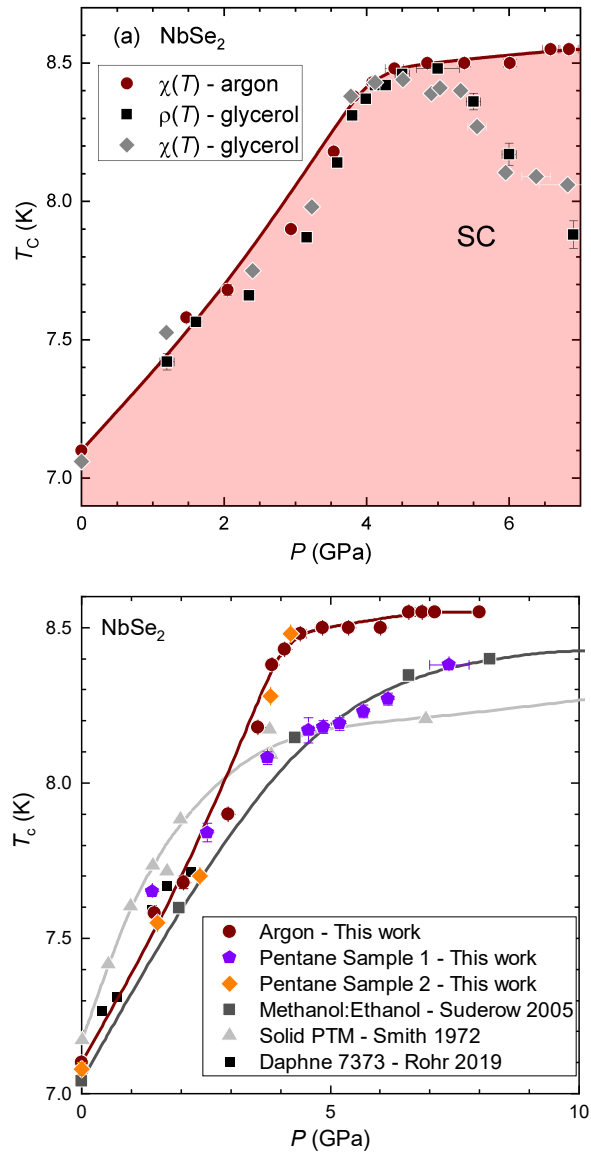

FIG. S3. Superconductivity in 2H-NbSe$_2$. (a) Comparison of measurements using argon and glycerol as pressure media. (b) Comparison of our measurements with previous studies beyond the CDW critical pressure [6, 7].

behaviour of $T_c(P)$: For all samples showing the clear kink in $T_c(P)$ (as observed with argon PTM), transitions in $\chi(T)$ are very sharp (see Fig. 3 of the main text and Fig. S4(a) and (c)). For samples with a reduced $T_c$ and a smooth rise in $T_c(P)$, transitions in $\chi(T)$ are much broader (Fig. S4(b)). We note that the broad transitions in glycerol above the solidification pressure (Fig. S4(a) for $P \gtrsim 5.5$ GPa) are attributed to the uniaxial component of the pressure as discussed above. With two samples measured in pentane:isopentane following different behaviour in $T_c(P)$ and in $\chi(T)$ we identify the differences to arise from sample preparation. Indeed, sample 2 and our sample in argon have been screened for sharp

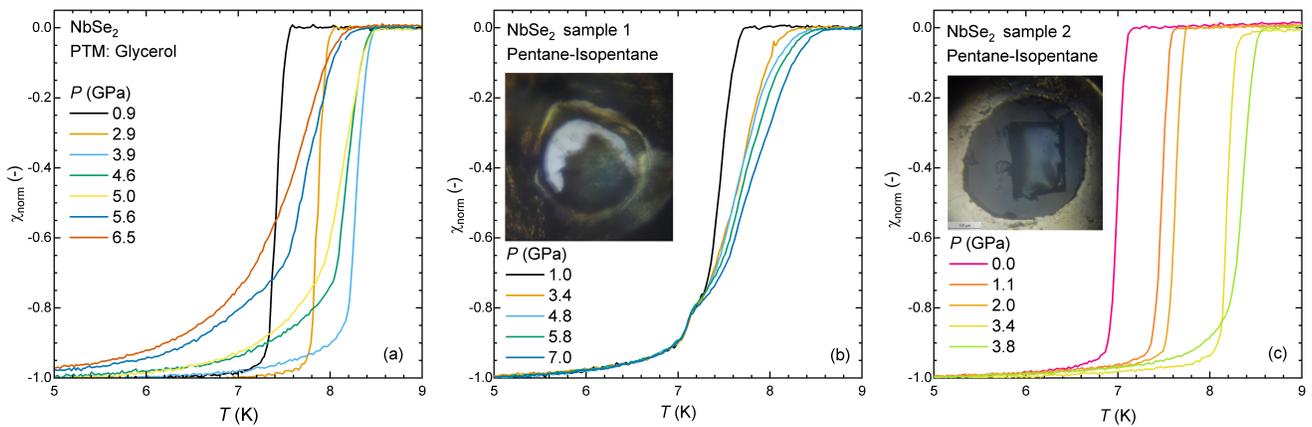

FIG. S4. Superconducting transition of 2H-NbSe$_2$ in a pressure cell using glycerol as the pressure transmitting medium (a). After subtraction of the background as described in the Methods section of the main manuscript, $\chi$ has been normalised $\chi_{\mathrm{norm}}(T) = (\chi(T) - \chi(9\,\mathrm{K}))/(\chi(9\,\mathrm{K}) - \chi(4\,\mathrm{K}))$. Measurements were done during heating up in $\mu_0 H = 0.5\,\mathrm{mT}$ after zero-field cooling. Pictures in (b) and (c) show the sample inside the pressure cell at high pressures.

transitions at ambient pressure before the study.

With best hydrostatic conditions and sharp transitions in $\chi(T)$ correlated to a sharp kink in $T_{\mathrm{c}}(P)$ at the critical pressure of the CDW we conclude that the behaviour observed with argon as a pressure medium reveals the intrinsic behaviour of 2H-NbSe$_2$ under pressure.

## S III.   HIGH-PRESSURE RESISTIVITY MEASUREMENTS

Fig. S5 shows a zero-pressure resistivity trace and its temperature derivative for a sample from the same batch like the high-pressure measurements. The CDW transition is observed at 33.4 K.

Fig. S6 shows the signature of the CDW in the electrical resistivity in our high-pressure measurements. We trace $T_{\mathrm{CDW}}(P)$ as the minimum in $\mathrm{d}\rho/\mathrm{d}T$ as shown in the inset of Fig. S6.

The upper critical field $H_{\mathrm{c2}}(T)$ has been extracted from temperature sweeps measuring $T_{\mathrm{c}}$ at a fixed field. At each field $T_{\mathrm{c}}$ was determined as the temperature where the resistivity reaches 10 % of the normal state value (see Fig. S7).

We use the Ginzburg-Landau equation

$$\left.\frac{\mathrm{d}\mu_0 H_{\mathrm{c2}}}{\mathrm{d}T}\right|_{T_{\mathrm{c}}} = -\frac{2.83\pi^2 k_{\mathrm{B}}^2}{e\hbar}\frac{T_{\mathrm{c}}}{v_{\mathrm{F}}^2} \qquad \text{(S1)}$$

to extract the Fermi velocity $v_{\mathrm{F}}$ from the slope of the critical field. Our measurements show a linear slope over a similar large temperature range as previous studies [9].

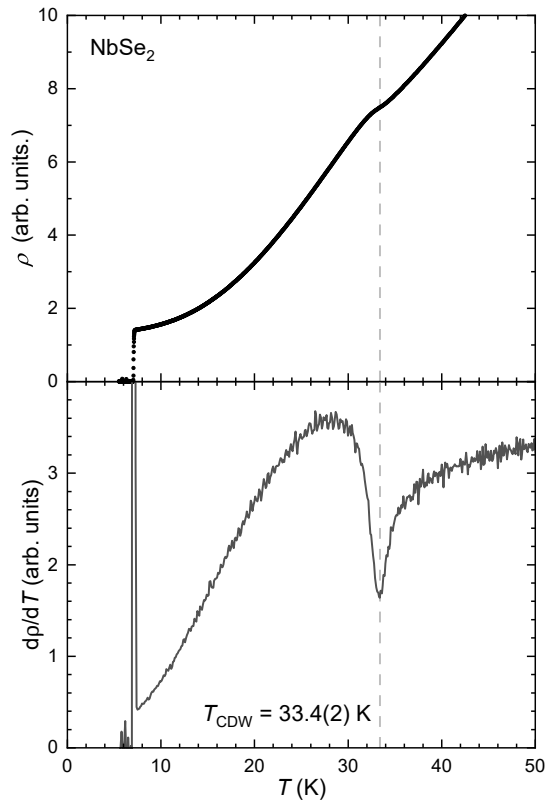

FIG. S5. Zero-pressure resistivity measurements on a sample of 2H-NbSe$_2$ from the same batch as used for the high-pressure resistivity measurements.

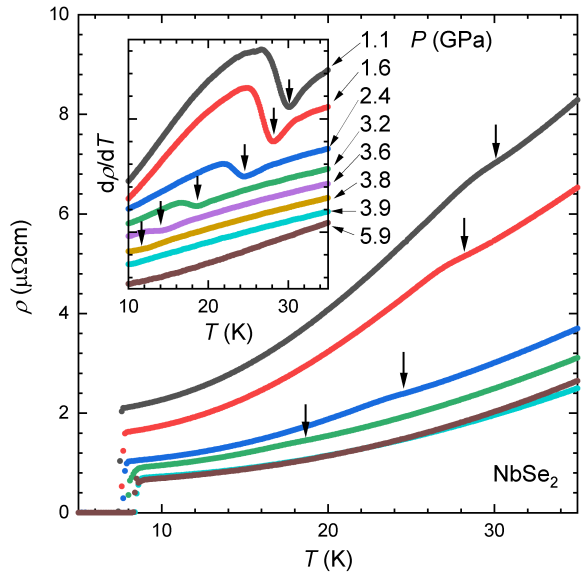

FIG. S6. Electrical resistivity at selected pressures. Inset shows the derivative d$\rho$/d$T$. Data in the inset have been offset for clarity. Arrows indicated the minimum in d$\rho$/d$T$ on both panels.

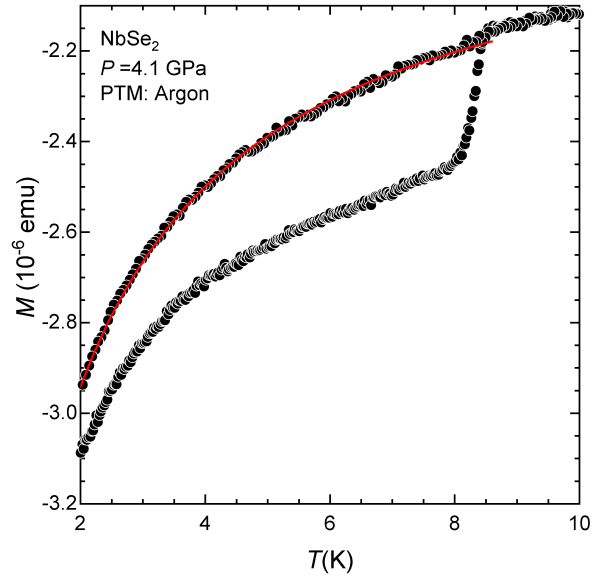

FIG. S8. Magnetic moment of the pressure cell with 2H-NbSe$_2$ sample on warming after zero-field cooling and on field cooling in $\mu_0 H = 0.5$ mT. The red line shows the Curie-type fit of the background.

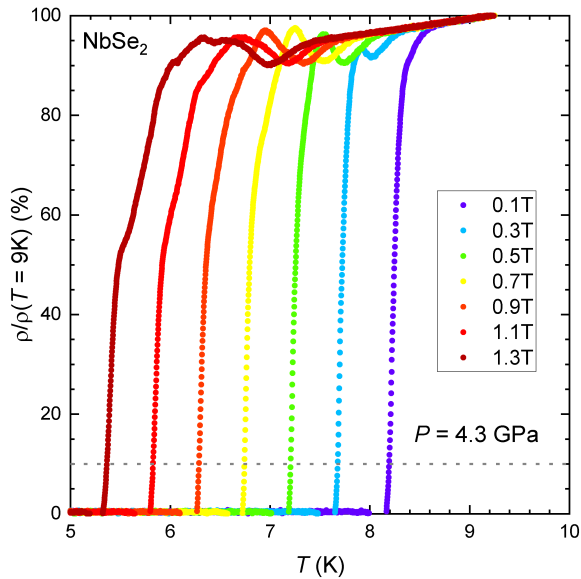

FIG. S7. Resistivity of 2H-NbSe$_2$ normalised to the normal state resistivity at $T = 9$ K. Dotted line indicates 10 % which was used to determine the critical field $H_{c2}(T)$ presented in Fig. 4 of the main text.

## S IV. BACKGROUND SUBTRACTION IN MAGNETIC MEASUREMENTS

A pressure cell mirror symmetric about the sample position was used. Thus, the MPMS software is able to reliably fit a dipole function with the amplitude giving the total magnetic moment of the sample and pressure cell.

We remove the background contribution arising from the pressure cell by subtracting a Curie-Weiss type contribution fitted to the field-cooled measurement as illustrated in Fig. S8.

The demagnetisation factor, $D$, of the sample was calculated using the rectangular prism approximation [10]. The magnetic susceptibility, $\chi$, is calculated from the sample magnetic moment $m_s$, sample volume $V_s$, and the applied static magnetic field $H$ as

$$\chi = \frac{m_s}{V_s H}(1 - D). \tag{S2}$$

Small variations in $\chi$ at low temperatures are associated with uncertainty of the sample position relative to the SQUID pick-up coils.

## S V. CALCULATION OF THE CHARGE-DENSITY-WAVE TRANSITION TEMPERATURE

We employed diagrammatic expansions based on the Random Phase Approximation (RPA), assuming the CDW to develop from a structured electron-phonon coupling dependent on both the ingoing and outgoing electron momenta and the orbital content of the bands. This model, which has as its only free parameter the overall magnitude of the electron-phonon coupling (fixed by $T_{CDW}(P = 0)$), has previously been shown to agree well with the full range of experimental observations on the charge ordered state in 2H-NbSe$_2$.

The Random Phase Approximation provides the following expression for the softening of the bare (high-temperature) phonon mode $\Omega_0(q)$ as a function of momentum transfer $\mathbf{q}$ as the temperature decreases towards the CDW phase transition at $T_{\text{CDW}}$:

$$\Omega(q,T)^2 = \Omega_0(q)(\Omega_0(q) - D_2(q,T))$$

where $D_2(q,T)$ is the generalized susceptibility to CDW formation [2, 11, 12]. $D_2(q,T)$ is the Lindhard function convolved with the square of the electron-phonon coupling. As temperature decreases $D_2(q,T)$ increases until, at $T_{\text{CDW}}$ and wavevector $Q_{\text{CDW}} \approx 0.986 \cdot \frac{2}{3}\Gamma M$, $D_2(Q_{\text{CDW}}, T_{\text{CDW}}) = \Omega_0(Q_{\text{CDW}})$, and $\Omega$ softens to zero. A CDW with wavevector $Q_{\text{CDW}}$ results. The one free parameter in the model, the magnitude of the electron-phonon coupling, is set to give the measured value of $T_{\text{CDW}} = 33.4\,\text{K}$ at $P = 0$.

Generally, the effect of increased pressure is to increase the frequency of phonons. Thus, we model high pressures as an increase of $\Omega_0$, the frequency of the longitudinal acoustic mode from which the CDW develops in 2H-NbSe$_2$. We find that the momentum transfer $q$ at which $D_2(q,T)$ peaks is largely independent of temperature down to $T = 0$, in agreement with temperature and pressure-dependent X-ray diffraction results [13]. Thus, we can obtain the pressure dependence of $T_{\text{CDW}}$ from $D_2(Q_{\text{CDW}}, T = 0)$ using the ambient-pressure $Q_{\text{CDW}}$.

As a consequence of the increase of $\Omega_0(Q_{\text{CDW}})$ at higher pressures, a larger $D_2(Q_{\text{CDW}}, T)$ is required to reach $\Omega = 0$ necessary to achieve the CDW transition. As $D_2$ increases with decreasing temperature, this corresponds to a decrease in $T_{\text{CDW}}$ with pressure. In our model calculations, we analyse the isothermal behaviour: For a fixed $D_2(Q_{\text{CDW}}, T)$, i.e. for a fixed temperature, we find the pressure $P_{\text{CDW}}(T)$ at which $\Omega(q,T)^2 = 0$ corresponding to the boundary of the CDW phase. The value of $D_2(Q_{\text{CDW}}, T = 0)$ sets the maximum bare phonon energy (and therefore pressure) from which a CDW can develop. From this we extract a pressure scaling factor to fit the experimentally observed phase boundary. This scaling corresponds to a rate of stiffening of the bare phonon mode. Finally, by inverting the relation we obtain $T_{\text{CDW}}(\text{P})$, the phase boundary plotted in Fig. 2 of the main manuscript.

## S VI.   CALCULATION OF THE SUPERCONDUCTIVITY TRANSITION TEMPERATURE

Our calculation of $T_c$ as a function of pressure is based on the change of the density of states (DoS) at the Fermi level, $g_{E_{\text{F}}}(\Delta_{\text{CDW}})$, as a function of the CDW gap magnitude $\Delta_{\text{CDW}}$.

The total DoS $g_{E_{\text{F}}}(\Delta_{\text{CDW}})$ is calculated from the contributions of the two Nb $d_{3z^2-r^2}$ orbitals $g_{E_{\text{F}}}^{\text{Nb}}(\Delta_{\text{CDW}})$, captured by our model of the Fermi surface developed in Refs. [2, 11, 12]. We add a contribution $g_{E_{\text{F}}}^{\text{Se}}(\Delta_{\text{CDW}})$

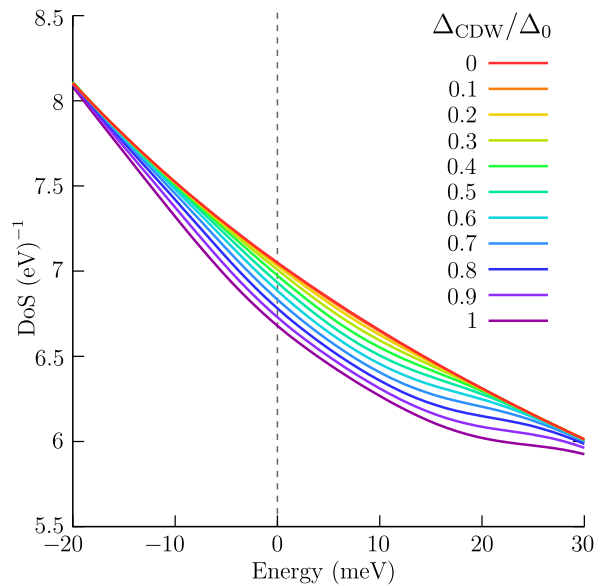

FIG. S9. Density of states $g^{\text{Nb}}(E)$ of the Nb orbitals as a function of energy for different values of the gap, normalised to the gap value at zero pressure. The energy is given relative to $E_{\text{F}}$.

from the selenium band such that the total matches the Sommerfeld coefficient [14].

The DoS of the Nb orbitals was calculated as the sum over the Brillouin zone of the spectral function $A(E,\mathbf{k})$:

$$g^{\text{Nb}} = \frac{1}{N}\sum_{\mathbf{k} \in BZ} A(E,\mathbf{k}) = -\frac{1}{\pi N}\mathfrak{Im}\sum_{\mathbf{k} \in BZ} G(E,\mathbf{k})$$

where $G(E,\mathbf{k})$ is the retarded electronic Green's function at energy $E$ and wavevector $\mathbf{k}$. We calculated the Green's function, including the CDW gap, using the Nambu-Gor'kov method [2]. For the wavevector dependence of the CDW gap we solved for the gap self-consistently at six high-symmetry points across the Brillouin zone, and used the results to create a six-parameter tight-binding fit. This calculation was previously shown to give a good match to scanning tunneling spectroscopy measurements of $g(E)$ over a range of energies around $E_{\text{F}}$ [15]. Fig. S9 shows this result. Note that the CDW gap is centred 16 meV above $E_{\text{F}}$; nevertheless, it is $g_{E_{\text{F}}}$ which is relevant for the formation of superconductivity.

We simulate the pressure dependence of $T_c$ assuming that $\Delta_{\text{CDW}}$ varies from the zero-pressure value $\Delta_0 = 12\,\text{meV}$ down to zero. We obtain the DoS of the Nb orbitals $g_{E_{\text{F}}}^{\text{Nb}}(\Delta_{\text{CDW}})$ as shown in Fig. S10. We assume a BCS temperature dependence of $\Delta_{\text{CDW}}(T)$ at a given pressure to obtain $g_{E_{\text{F}}}^{\text{Nb}}(\Delta_{\text{CDW}})$ at the superconducting transition temperature self consistently.

We relate the CDW gap to pressure by scaling to the fitted $T_{\text{CDW}}(P)$ in Fig. 2(a):

$$P = P_{\text{CDW}}(T=0)\left[1 - \left(\frac{\Delta_{\text{CDW}}}{\Delta_0}\right)^{1/n}\right] \quad \text{(S3)}$$

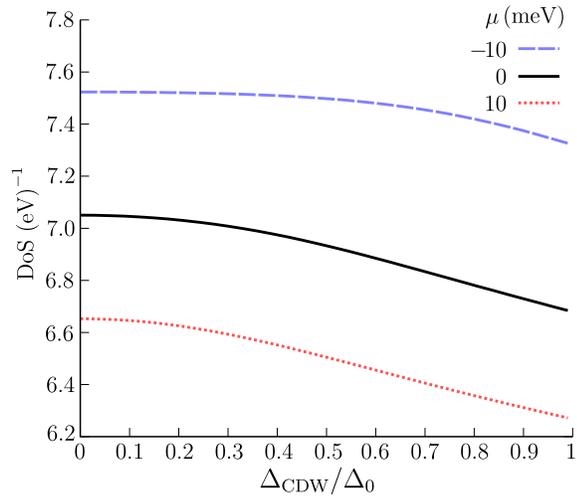

FIG. S10. Density of states $g_{E_F}(\Delta_{CDW})$ at the Fermi level as a function of the CDW gap size $\Delta_{CDW}$.

where $n$ and $P_{CDW}(T=0)$ are the exponent and critical pressure from the fit to the phase boundary (solid line in Fig. 2).

In order for the total density of states $g_{E_F}$ to be consistent with the Sommerfeld coefficient, we add a constant value of $0.0013\,\mathrm{meV}^{-1}$ which is associated with the DoS from the Se-orbitals.

We use the BCS expression to calculate the transition temperature of the superconducting state

$$T_c = 1.14\Theta \exp\left(-\frac{1}{g_{E_F}V}\right)$$

where the coupling constant $V = 0.035$ was obtained to fit the zero-pressure $T_c$ using the zero-pressure $g_{E_F}$ (which is consistent with the experimentally determined DoS from the Sommerfeld coefficient).



\end{document}